\documentclass{aa}
\usepackage{amsmath,graphicx,amssymb}
\usepackage{txfonts}
\usepackage{natbib}
\bibpunct{(}{)}{;}{a}{}{,}
\def\urltilda{\kern -.15em\lower .7ex\hbox{\~{}}\kern .04em}
\newcommand\hvezda{\object{HR\,7355}}
\newcommand{\zav}[1]{\left(#1\right)}
\newcommand{\hzav}[1]{\left[#1\right]}

\newlength\staretab

\makeatletter
\def\sgn{\mathop{\operator@font sgn}\nolimits}
\makeatother

\begin{document}

\title{HR\,7355 -- another rapidly braking He-strong CP star?}
\titlerunning{HR\,7355 -- another rapidly braking CP star?}

\author{Z. Mikul\'a\v sek\inst{1,5}\and J. Krti\v{c}ka\inst{1}
\and G. W. Henry\inst{2}\and S. N. de Villiers\inst{3}\and E.
Paunzen\inst{4}\and M. Zejda\inst{1} }
\authorrunning{Z.~Mikul\'a\v sek et al.}
\offprints{Zden\v ek~Mikul\'a\v sek,\\
\email{mikulas@physics.muni.cz}}

\institute{Department of Theoretical Physics and Astrophysics,
           Masaryk University, Brno, Czech Republic
        \and
            Center of Excellence in Information Systems, Tennessee
            State University, Nashville, Tennessee, U.S.A
        \and
            Private Observatory, 61 Dick Burton Road, Plumstead,
            Cape Town, South Africa
        \and
            Institute for Astronomy of the University of Vienna,
            Vienna, Austria
        \and
            Observatory and Planetarium of J. Palisa, V\v SB --
            Technical University, Ostrava, Czech Republic}

\date{Received 12 December 2009/ Accepted 5 February 2010}

\abstract{Strong meridional mixing induced by rapid rotation is one
reason why all hot main-sequence stars are not chemically peculiar.
However, the finding that the He-strong CP star HR~7355 is a rapid
rotator complicates this concept.} {Our goal is to explain the
observed behaviour of HR~7355 based on period analysis of all
available photometry.}{Over two years, we acquired 114 new
\textit{BV} observations of \hvezda\ at observatories in Arizona,
U.S.A and Cape Town, South Africa. We performed period analyses of
the new observations along with new analyses of 732 archival
measurements from the Hipparcos and ASAS projects.} {We find that
the light curves of \hvezda\ in various filters are quite similar,
with amplitudes 0.035(4), 0.036(4), and 0.038(3) mag in $B$,
\textit{Hp}, and $V$, respectively. The light curves are
double-peaked, with unevenly deep minima. We substantially refine
the rotational period to be $P=0\fd5214410(4)$, indicating that
\hvezda\ is the most rapidly rotating CP star known. Our period
analyses reveal a possible lengthening of the rotational period with
$\dot{P}/P=2.4(8)\times 10^{-6}\,\rm{yr}^{-1}$.} {We conclude that
the shape and amplitude of \hvezda\ light curves are typical of
magnetic He-strong CP stars, for which light variations are the
result of photospheric spots on the surface of a rotating star. We
hypothesise that the light variations are caused mainly by an uneven
distribution of overabundant helium on the star's surface. We
briefly describe and discuss the cause of the rapid rotational
braking of the star.}


\maketitle

\section{Introduction}

Chemically peculiar (CP) stars are an important class of stars that
occupy the upper main sequence, where radiative diffusion and
gravitational settling result in atmospheric chemical abundances
that differ remarkably from the Sun's. According to their patterns
of chemical anomalies, CP stars are classified into several
subclasses that also follow a temperature sequence. Among them,
SrCrEu, Si, He-weak, and He-strong stars have strong global magnetic
fields. These ``magnetic'' CP (mCP) stars also exhibit synchronous
variability in their spectra and brightness with periods longer than
one half of a day. Their photometric amplitudes are a few hundredths
of magnitude.

mCP stars have inhomogeneous surface distributions of chemical
elements as determined from their rotationally modulated
spectral-line variability \citep[e.g.,][]{lufta,luftb}. The uneven
surface distribution of various elements, together with rotation,
has been presumed to be the main cause of these stars' light
variability. Line blanketing caused by overabundant elements, mainly
iron-group metals and \textit{b--f} transitions, may induce the flux
redistribution providing the mechanism for the light variability
\citep{lanko}. A strong magnetic field may also influence the light
curves \citep[e.g.,][]{malablaj}.

\citet{krt,eedra} used surface abundance maps to simulate
successfully the light curves of the He-strong star
\object{HD\,37776} and the Si star \object{HR\,7224}. They
demonstrated that the inhomogeneous surface distribution of silicon,
iron, and helium, along with the \textit{b--f} and \textit{b--b}
transitions of these elements, accounted for most of the light
variability in these CP stars.

The light curves of magnetic CP stars are stable on a timescale of
decades or more and so indicate persistency of their spectroscopic
and photometric spots. We can observe period changes in their light
curves in only a few cases \citep[][and references therein]{mikbra}.

It is generally expected that the meridional currents induced by the
rapid rotation of hot MS stars are able to erase the effects of the
slow diffusion of chemical elements and thus prevent the formation
of CP abundance anomalies. We study therefore the light variability
of \hvezda, which is one of the most rapidly rotating mCP stars.

\section{The star}

\hvezda\ (HD\,182180, HIP\,95480) is a bright ($V=6.02$ mag) but
poorly studied southern B2V star. It is known to be a rapid rotator,
\citet{oks} evaluating its projected equatorial velocity to be
$v\,\sin\,i=(300\pm15)\,\mathrm{km\,s}^{-1}$. The Hipparcos catalog
\citep{ESA} classified the star as an ``unsolved variable'', but
\citet{koen} reanalysed the Hipparcos photometry and found a
frequency of variation of $1/0\fd26072$.

The first detailed study of \hvezda\ was performed by \citet{riv},
who describe the star as a helium-strong CP star with Balmer
emission. They argue that the true period, corresponding to the
star's rotational period, is twice that derived by \citet{koen}:
$P=0\fd521428(6)$. This suggests that \hvezda\ is the most rapidly
rotating He-strong CP star known.

In the present study, we analyse our new and archival photometry of
\hvezda\ to determine the star's periodicity and light curve shape
at various epochs between 1990 and 2009.

\subsection{Parameters of the star}

We estimate the mass and age of \hvezda\ using the revised version
of the Hipparcos catalogue \citep{leeuwen}, the Geneva and
Str{\"o}mgren photometry (from
GCPD\footnote[1]{http://obswww.unige.ch/gcpd}), and the isochrones
of the Padova group \citep{marigo}.

The Geneva colours \citep{cramer} are consistent with a B2\,V star.
The $Z$ index shows no peculiarity beyond what we would expect for a
classical CP4 star \citep{paun}. From the various Geneva indices and
the available Str{\"o}mgren photometry, we deduce a reddening of
$E(b-y)\,=\,0.05$ mag and an effective temperature of 18\,000\,K.
The photometric estimates for the absolute magnitude, $M_V$, vary
between $-1.75$ and $-1.95$\,mag. The parallax gives a distance of
273(26)\,pc and an absolute magnitude of $M_{V}=-1.38(0.20)$\,mag
(assuming $V$\,=\,6.02\,mag). This result is fainter than those
estimated from the photometric indices, which may be a result of the
star's rapid rotation.

We estimate the age of \hvezda\ to be between 15 and 25\,Myr, as
derived from the parallax-based and the photometry-based values of
$M_V$, respectively. We note that \citet{westin} lists $\log
t=7.379$ (23.9\,Myr) for \hvezda. Consequently, \hvezda\ appears to
be near the middle of its main-sequence lifetime. We derive a mass
of 6.3(0.3)~M$_{\sun}$ from the parallax-based $M_V$. The
photometry-based absolute magnitude would result in a higher mass.

From its distance and the Galactic coordinates, we derive the
Galactic [X,Y] distances/coordinates of \hvezda\ to be
$[+254,+48]$~pc. Comparing these values and their uncertainties with
the borders of the Scorpius-Centaurus association \citep{prema}, we
find that \hvezda\ is probably beyond the outer edge of the
association.  The mean proper motion of the association is
$[-25,-10]$~mas~yr$^{-1}$ \citep{zeeuw}, compared to
$[+12,-15]$~mas~yr$^{-1}$ for \hvezda. Given its motion direction
and apparent age, we suggest that \hvezda\ does not originate in the
Scorpius-Centaurus association.

\section{Photometric data of \hvezda}

\subsection{New photometry}

Most of our new photometry of \hvezda\ was acquired with the T3
0.4\,m automatic photoelectric telescope (APT) at Fairborn
Observatory in southern Arizona during 2008 May and June. The T3 APT
uses a temperature-stabilised EMI 9924B photomultiplier tube to
measure photon count rates sequentially through $B$ and $V$ filters.
For additional details about the collection and reduction of the APT
data, see cf. \citet{mikbra} and references therein. The comparison
and check star used by the APT were HD\,179520 and HD\,181240,
respectively. The southern declination of \hvezda\
($\delta=-28^{\circ}$) causes airmass values of $\sim2$ for all of
our APT measurements. The standard deviation of the difference
between variable and comparison stars is $\sim0.01$ mag, twice the
typical scatter with this telescope. Thus we acquired 43 $B$ and 38
$V$ measurements.

We also obtained seven continuous hours of $BV$ photometry with a
photoelectric photometer attached to a 0.28~m reflector on the night
of 2009 July 26 UT at the private observatory of one of us (SNdV) in
Cape Town, South Africa. We acquired 17 and 16 measurements in $V$
and $B$. The comparison and check stars were the same as used by the
APT; the scatter in the observations was 0.01 mag. When combined
with the APT measurements, these continuous single-night
observations helped us to minimise the aliasing inherent in
single-sight data sets.

The newly acquired photometric data are available through
SIMBAD\footnote{http://cdsweb.u-strasbg.fr/cgi-bin/qcat?J/A+A/} or
the {\it On-line database of photometric observations of mCP
stars}\footnote{http://astro.physics.muni.cz/mcpod}.

\subsection{Archival photometry}

The Hipparcos photometry \citep{ESA} of \hvezda\ consists of 57
measurements in $B_{\mathrm{T}}$ (scatter $\sigma$ of 23\,mmag), 45
in \emph{Hp} ($\sigma=9$\,mmag), and 56 in $V_{\mathrm{T}}$
($\sigma=28$ mmag). The data cover the period from 1990 March to
1993 March.

The continuously updated archive of the All Sky Automated Survey
(ASAS) \citep{pojm} is a useful source of photometric observations
of stars near the equator. Unfortunately, the precision for
relatively bright stars is poor -- the data have significant
non-Gaussian scatter with plenty of outliers. Furthermore, the data
contain a long-term trend with a range of 0.05~mag.

Nonetheless, the ASAS data are the only photometric data of \hvezda\
available in the period between the end of the Hipparcos mission
(1993) and the start of our own photometry (2008). We extracted only
the individual $V$ measurements with a quality of ``A'' or ``B'' and
removed the long-term trend. This resulted in an ASAS data set of
574 $V$ observations covering the time interval from 2001 February
to 2009 October. These data have a scatter of 50\,mmag.

Therefore, we consider in our period analysis a total of 846
measurements spread over 20 years.

\section{The ephemeris}

\subsection{The period}

\begin{figure}[t]
\centering \resizebox{0.85\hsize}{!}{\includegraphics{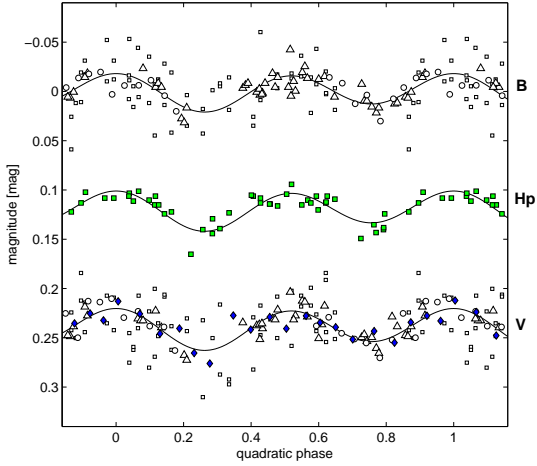}}
\caption{\textit{BHpV} light curves plotted versus the quadratic
phase. $\Box$ -- Hipparcos, $\circ$ -- Cape Town, $\triangle$ --
APT, $\blacklozenge$ -- ASAS (each point represents the mean of 30
adjacent ASAS measurements).} \label{krivky}
\end{figure}

The periodogram of all available \hvezda\ photometry displays only
two prominent peaks -- one at a period of $0\fd52144$ and the second
at $0\fd26072$ -- that agree well with the previous determinations
of \citet{koen} and \citet{riv}. The shorter period is the
consequence of the double-peaked nature of the light curve (see Fig.
\ref{krivky}). This $0\fd26072$ period is shorter than the critical
period of centrifugal disruption of main-sequence stars of the same
spectral type and so cannot represent \hvezda's rotation period,
which we take to be $0\fd52144$. The longer period is also
compatible with the period of the well documented spectral and
spectropolarimetric variations \citep{oks,riv2}.

\subsection{The light variations model}

\begin{figure}[t]
\centering \resizebox{0.86\hsize}{!}{\includegraphics{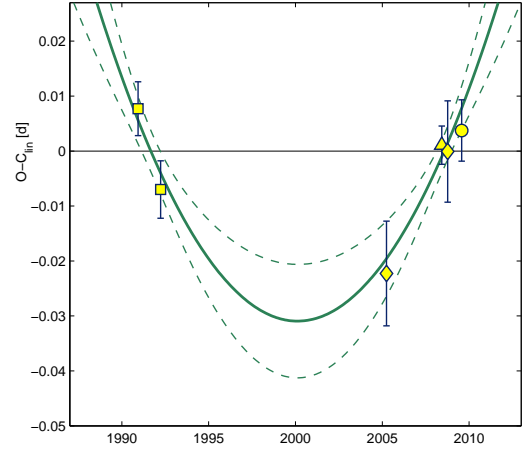}}
\caption{Time variation in mutual phase shifts of observed light
curves in days computed with a linear ephemeris. $\Box$ --
Hipparcos, $\diamondsuit$ -- ASAS, $\triangle$ -- Arizona, and
$\circ$ -- Cape Town. The Hipparcos and ASAS data sets are divided
into two pieces with equal numbers of observations and plotted
separately. This division was made for only illustrative reasons and
did not affect the period analysis in any way. The quadratic time
function is denoted by the solid line, while the
1\,$\sigma$-uncertainties are plotted with dashed lines. }\label{OC}
\end{figure}

The double-waved light curves in $B$, \textit{Hp}, and $V$ are
similar (see Fig.\,\ref{krivky}), differing only slightly in their
effective amplitudes $A_c$ \citep[for the definition see][]{mikAN},
where the subscript $c$ denotes the filter colour. Light-curve
magnitudes can then be expressed as
\begin{equation}\label{krivka}
 m_{cj}(t)\simeq\overline{m}_{cj}+\textstyle{\frac{1}{2}}\,A_c
 F(\vartheta),
\end{equation}
where $m_{cj}(t)$ is the magnitude in colour $c$ observed by $j$-th
observer, $\overline{m}_{cj}$ is the mean magnitude, which can be
variable over the long term. This was the case for ASAS
measurements, where we had to assume a cubic trend. The function
$F(\vartheta)$ is the simplest normalised periodic function that
represents the observed photometric variations of \hvezda\ in
detail. The phase of maximum brightness is defined to be 0.0, and
the effective amplitude is defined to be 1.0. The function, being
the sum of three terms, is described by two parameters
$\beta_1$\,and $\beta_2$, where $\beta_1$ quantifies the difference
in depths of the primary and the secondary minima and $\beta_2$
expresses any asymmetry in the light curve
\begin{eqnarray}\label{fce}
F(\vartheta,\, \beta_1,\,\beta_2)=\beta_1\,\cos(2\,\pi\,\vartheta)+
\sqrt{1\!-\!\beta_1^2\!-\!\beta_2^2}\ \cos(4\,\pi\,\vartheta)\nonumber \\
+\,\beta_2\hzav{\textstyle{\frac{2}{\sqrt{5}}}\,\sin(2\,\pi\,\vartheta)
-\textstyle{\frac{1}{\sqrt{5}}}\sin(4\,\pi\, \vartheta)},
\end{eqnarray}
where $\vartheta$ is the \textit{phase function} of time described
in \citet{mikbra}. The O-C diagram (see Fig. \ref{OC}) indicates
that the function may be parabolic; thus, we assumed it to have the
form
\begin{equation}
\vartheta \cong
\vartheta_1-\textstyle{\frac{1}{2}}\dot{P}\zav{\vartheta_1-\alpha_1}
\zav{\vartheta_1-\alpha_2};\quad \vartheta_1=\zav{t-M_{01}}/P_1,
\label{inverse}
\end{equation}
where $M_{01}$ is the maximum nearest the weighted center of the
observations, $P_1$ is the period of the linear fit, $\dot{P}$ is
the time derivative of the period (assumed to be constant here), and
$\alpha_1$ and $\alpha_2$ are constants determined by the data
distribution in time so that the first and second terms in
Eq.\,\eqref{inverse} are orthogonal, such that
$\alpha_{1,2}=\Theta_1 \pm
\sqrt{\Theta_1^2+\overline{\vartheta_1^2}}$, where $\Theta_1=
\overline{\vartheta_1^3}/({2\,\overline{\vartheta_1^2}})$.

All 19 model parameters were computed simultaneously by a weighted
non-linear LSM regression applied to the complete observational
material without any artificial divisions.

\subsection{Results}

We determined the model parameters $\alpha_1=4420,\ \alpha_2=-7416$,
$M_{01}=2\,452\,367.1221(24),\ P_1=0\fd 5214410(4),\
\dot{P}=3.4(1.1)\times 10^{-9}=0.107(35)$ \,s\,y$^{-1}$, and
$\dot{P}/P_1=2.4(8)\times 10^{-6}\,\mathrm {yr}^{-1}$. Time of the
primary maximum being given by
\begin{eqnarray}\label{number}
\mathrm{HJD}_{\mathrm{maxI}}(E)\cong 2\,452\,367.1221(24)+0\fd
5214410(4)\times E\nonumber \\
+8\fd8(2\fd9)\times10^{-10}\ (E+7416)\,(E-4420),
\end{eqnarray}
where $E$ is an integer. The effective amplitudes $A_c$ are almost
identical and equal to 0.035(4), 0.036(4), and 0.038(3) mag in
\textit{B, Hp}, and $V$, respectively, $\beta_1=0.08(6)$, and
$\beta_2=-0.26(6)$.

The reliability of the LSM procedure results depends critically on
the appropriateness of the chosen model for the description of the
reality -- in our case light variations over the past twenty years.
Inspecting Fig. \ref{krivky}, we can conclude that the model of
light curves described by Eqs.\,\eqref{krivka} and \eqref{fce} is
fully adequate.

Adequacy of the quadratic model used for the phase function
Eq.\,\eqref{inverse} can be tested by changes in mutual phase shifts
of observed light curves. The shift (O-C)$_k$ can be evaluated for
any subset $k$ of observational data by the relation:
(O-C)$_k=-\overline{\Delta
m_{k}\,\dot{m}_{k}}/\overline{\dot{m}^2_{k}}$, where $\Delta m_{k}$
are deviations of observed magnitudes from the fit, and
$\dot{m}_{k}$ are time derivatives of the fitted function
\citep{mikwol}. From Fig.\,\ref{OC} containing shifts of 6
observational data subsets with respect to the linear ephemeris, we
conclude that our quadratic model of the phase function in
Eq.\,\eqref{inverse} is acceptable.

To be sure that non-zero $\dot P$ is not a mere conjunction of the
non-Gaussian noise in ASAS data, we treated the observational set
excluding all the ASAS data. Even in this case, we obtained $\dot
P=2.8(1.6)\times 10^{-9}$, supporting the reality of the period
lengthening.

\section{Discussion}

\subsection{\hvezda\ -- the most rapidly rotating CP star}

The $P_1=0\fd52144$ rotational period of \hvezda\ is among the
shortest known to date for CP stars: HD\,164429
\citep[$P=0\fd51899$,][]{adel}; HD\,124224\,=\,CU\,Vir
\citep[$P=0\fd 52070$,][]{sok}; HD\,92385
\citep[$P=0\fd54909$,][]{ESA}. Among these fast rotating CP stars,
\hvezda\ is the hottest, most massive and, consequently, the largest
thus, we conclude that \hvezda\ has the largest equatorial
rotational velocity ($V_{\mathrm{eq}}\approx 370 \pm 80 \ \mathrm
{km\,s}^{-1}$ using the parallax-based $M_V$ for a B2\,V star and
the period) and also the highest ratio of equatorial velocity to
critical equatorial velocity $(0.75\pm0.25)$ among all known CP
stars. Therefore, we consider \hvezda\ to be the most rapid rotator
of all known chemically peculiar stars. Because of its rapid
rotation, it can serve as an important benchmark for theories
describing the influence of rotational mixing on chemical
peculiarity.

\subsection{Nature of the light variation}

Our analysis of the $B$, \textit{Hp}, and $V$ light curves
demonstrates that all three passbands have the same effective
amplitude. This suggests that the variability mechanism at optical
wavelengths could be unique \citep{miksim}. We suggest that these
light variations are caused by the uneven distribution of optically
active, overabundant elements on the surface of the star.
Unfortunately, we do not have maps of the abundance distribution or
spectrograms suitable for their creation.

We consider the possibility that the broad-band optical light
variability in \hvezda\ arises from variations in the spectral lines
of $\ion{He}{i}$, $\ion{Si}{iii}$, and $\ion{C}{ii}$, as depicted
for \hvezda\ in Fig.~1 of \citet{riv}.  They compare the line
profiles of two spectra obtained at quadrature (HJD
$=2\,451\,385.507,\ \vartheta=-1882.445(20)$ and HJD
$=2\,453\,191.879,\ \vartheta=1581.731(15)$, as defined for
Eq.\,\eqref{inverse}); their first spectrum was thus taken shortly
after secondary light maximum, while the second was acquired at the
secondary light minimum. The only overabundant element whose line
intensity reaches maximum in the optical is helium \citep[see the
relevant analysis of He-strong HD\,37776 in][]{krt}. Lines of the
other ions mentioned above are relatively weak and nearly constant.
The photometric effect of the weak emission in H$\alpha$ is also
negligible.

We modelled the light curves of \hvezda\ with the code described by
\citet{krt} assuming the inclination angle of
$\mathit{i}=54^\circ\!\pm18^\circ$ \citep[$v\,\sin\,i=300\pm\!15$
km\,s$^{-1}$,][]{oks} and found that two circular, helium-rich spots
with [He/H]=1.4 and radii $60^\circ$ on opposite hemispheres of
\hvezda\ reproduce the observed light variations in all three
passbands. We therefore conclude that light variations in \hvezda\
may be the result of the uneven distribution of helium.

This hypothesis should be tested with Doppler tomography, which we
plan to do in the near future.

\subsection{Rotational braking and its nature}

An increase in the rotational period with $\dot{P}/P_1=2.4(8)\times
10^{-6}\,\mathrm {yr}^{-1}$ is observed at the $3.1\,\sigma$ level
of certainty.  During the past 20 years, the period has increased by
2.1 s!  Unfortunately, no observations were taken in the interval
1994-2000. This adds some uncertainty to our quadratic fit of the
O-C residuals. If the period change were real, then \hvezda\ would
be the fourth known CP star displaying an increase in its rotational
period.

The star is similar in several aspects to the most rapidly braking
He-strong CP star HD\,37776 with a well-determined
$\overline{\dot{P}/P}=4.01(17)\times10^{-6}$\,year$^{-1}$
\citep{mikbra}. The period increase in \hvezda\ could also be
interpreted as a deceleration of the rotation of its \textit{surface
layers} due to momentum loss by a magnetically confined stellar wind
\citep[see also][]{ud}. However, a possible change in the rate of
period change for HD\,37776
($\ddot{P}=-29(13)\times10^{-13}$\,d$^{-1}$) suggests that the
process of the rotational braking need not be fully monotonic!

Both stars have strong magnetic fields; that of \hvezda\ is
approximately dipolar ranging in strength from -2\,kG to +2.5\,kG
\citep{oks}, while the field of HD\,37776 is dominated by a
quadruple component \citep{thola}. However, the main difference
between the stars discussed here is their age. While HD\,37776 is a
very young CP star with an age of around 1 Myr, \hvezda\ is about
twenty times older (see Sect. 2.1). Because the characteristic
braking time of \hvezda\ ($P/\dot{P}\approx 4 \times 10^{5}$\,yr) is
about fifty times shorter than its stellar age, it seems probable
that variations in rotational period are confined to the outer
layers of the star. The period variations could be cyclic with a
long interval of period increase followed by a rapid period
decrease. The period decrease may be caused by greater friction
between the surface and inner layers of the star caused by, e.g.,
reconnection events.  The length of this cycle can be roughly
estimated as the time when the surface layers lag behind the inner
part of the star by one revolution (i.e., $T_{\mathrm{cycle}}\sim
P\,\sqrt{\,2/\dot{P}}$). This produces estimates of 35 yr and 45 yr
for \hvezda\ and HD\,37776, respectively.  The hypothesis should be
tested a few decades hence with additional observations.

\section{Conclusions}

We have improved the accuracy of the rotational period of the most
rapidly rotating CP star \hvezda\ to $P=0\fd5214410(4)$ and
determined its rate of period change to be $\dot{P}/P=2.4(8)\times
10^{-6}\,\mathrm {yr}^{-1}$. We propose that the period variations
could be cyclic on a timescale of a few decades. Observed light
variations may be caused by the uneven surface distribution of
overabundant helium. \hvezda\ remains a very appealing target for
continued photometric and spectroscopic observations, as well as for
the modelling of its unusual behaviour.

\begin{acknowledgements}

Supports of grants GAAV IAA301630901 and MEB 060807/WTZ CZ-11/2008
are acknowledged.  GWH acknowledges long-term support from NASA,
NSF, Tennessee State University, and the state of Tennessee through
its Centers of Excellence program. EP acknowledges support by the
City of Vienna (Hochschuljubil{\"a}umsstiftung project:
H-1930/2008). We thank to dr. D.\,A. Bohlender for his inspiring
refereeing of the article.

\end{acknowledgements}

\end{document}